%% file: Article.tex
\documentclass[review,onefignum,onetabnum]{siamart171218}

\usepackage[colorinlistoftodos]{todonotes}
\usepackage{comment}
\usepackage[export]{adjustbox}
\usepackage{caption}
\usepackage{subcaption}
\usepackage{graphbox}

\input{ex_shared}

\ifpdf
\hypersetup{
  pdftitle={The unified framework for modeling credit cycles with Marshall-Walras price formation process and systemic risk assessment},
  pdfauthor={Kamil Fortuna and Janusz Szwabi\'{n}ski}
}
\fi

\myexternaldocument{ex_supplement}

\begin{document}

\maketitle

\begin{abstract}
Systemic risk is a rapidly developing area of research. Classical financial models often do not adequately reflect the phenomena of bubbles, crises, and transitions between them during credit cycles. To study very improbable events, systemic risk methodologies utilise advanced mathematical and computational tools, such as complex systems, chaos theory, and Monte Carlo simulations. In this paper, a relatively simple mathematical formalism is applied to provide a unified framework for modeling credit cycles and systemic risk assessment. The proposed model is analyzed in detail to assess whether it can reflect very different states of the economy. Basing on those results, measures of systemic risk are constructed to provide information regarding the stability of the system. The formalism is then applied to describe the full credit cycle with the explanation of causal relationships between the phases expressed in terms of parameters derived from real-world quantities. The framework can be naturally interpreted and understood with respect to different economic situations and easily incorporated into the analysis and decision-making process based on classical models, significantly enhancing their quality and flexibility.  
\end{abstract}



\section{Introduction}

``With the takeover of Credit Suisse by UBS, a solution has been found to secure financial
stability and protect the Swiss economy in this exceptional situation''. At the time of writing this article, on 19th March 2023, the Swiss National Bank announced a resolution to prevent the second largest Swiss bank from failure \cite{swiss_national_bank_swiss_2023}. To obtain an acquisition agreement, Swiss National Bank pledged a loan of up to \$104 billion, while the Swiss government issued for UBS a guarantee to assume losses up to \$9.6 billion \cite{bishop_ubs_nodate}.
There was one decisive reason for undertaking such measures: the mitigation of systemic risk.  

Systemic risk represents the possibility of collapse of the entire financial system, as opposed to collapses of individual parts and components \cite{ilin_uncertainty_2015}. The aim of this research area is to assess the risk of the financial system to secure its persistence and mitigate the consequences of its failure. Based on the results, banks around the world determine the dimension of the reserves held and, consequently, the interest rates and the amount of money in the economy \cite{basel_comittee_on_banking_supervision_basel_nodate}. Therefore, this research discipline is of fundamental importance for the functioning of modern societies and exhibits a significant impact on our daily life. This importance was confirmed by the recent 2022 Nobel Prize in Economics for the work of Bernanke, Diamond, and Dybvig on bank failures \cite{bernanke_financial_nodate, diamond_bank_1983}.

Systemic risk research utilises complex mathematical tools \cite{doldi_conditional_2021} to analyze very improbable events and their probable consequences \cite{jackson_systemic_2020}. Its main challenge is the fact that the vast majority of such events will never occur, so their probability and potential impact must be assessed on the basis of the very sparse set of real-world observations and simulations of unrealised outcomes \cite{taleb_black_2007}. The tools include network models of financial contagion and phase transitions \cite{elliott_financial_2014}, systems with feedback effects, self-fulfilling prophecies with possible switches between multiple equilibria \cite{diamond_bank_1983}, and complex computer calculations to simulate potential trajectories \cite{gill_high-performance_2021}.

Network models are used to define the measures of systemic risk \cite{bartesaghi_risk-dependent_2020,battiston_debtrank_2012} and the methods of their computation on real data \cite{bourgey_metamodel_2020,poledna_quantification_2021}. The measures are then utilised to propose methods of risk optimization in financial systems \cite{pichler_systemic_2021}. The solutions include the usage of financial instruments such as credit default swaps to effectively reorganize the network structure towards a safer one \cite{leduc_systemic_2017} and contingent convertible obligations to improve banks' solvency and survival ability \cite{feinstein_contingent_2023}. The developed methodologies are, in turn, used to improve decision making on the management of financial systems in the real world \cite{gill_high-performance_2021}. The potential of the research performed extends beyond the banking system and is applied to the analysis and valuation of the credit risk and contagion effects of reinsurance companies \cite{ceci_value_2020}, as well as to the identification of systematically important entities in the credit network of the entire state \cite{poledna_identifying_2018}.

On the other hand, classical financial models generally are not appropriate for reflecting improbable events \cite{taleb_black_2007}. They focus mainly on long stable periods with moderate variance throughout the time \cite{black_pricing_1973}. Transitions between different phases of credit cycles constitute real-world phenomena that classical advanced, stochastic, and mathematically complicated models struggle to explain \cite{greenspan_age_2007}. Among the known and popular models, GARCH can relatively better align with the real data due to the volatility clustering property \cite{bollerslev_generalized_1986}. It includes the feedback effect, because a higher  value of volatility in a period increases the probability of a higher value in the next one. Nevertheless, the model does not capture some asymetries characteristic for credit cycle data, where even long periods of stable but excessive growth can contribute to the risk accumulation and, therefore, increase the fragility of the system. Similarly, mean-reversion models \cite{lipe_mean_1994} do not explain the increase in fragility (risk of severe collapses) for stable growth periods and asymetries between stable growth, steep crisis, and relatively fast recovery. 

The aim of this paper is to provide a unified framework for modeling credit cycles and systemic risk assessment with the utilization of relatively simple, easy to use, and well-interpretable mathematical formalism. It is structured as follows. \Cref{sec:models_introduction} introduces the model with the mathematical analysis of its properties. \Cref{sec:phase_transitions} focuses on the topic of phase transitions and the derivation of system stability measures. \Cref{sec:cycle} provides an example of a credit cycle generated from the model with details on the real-world interpretation of parameters and their economic consequences. A summary of the research results is presented in \Cref{sec:summary}.

\section{Models and methods}
\label{sec:models_introduction}

\subsection{The Marshall-Walras equilibrium (MWE) model applied to credit market}

The mathematical model of loans, defaults, and interest rate dynamics with the autocatalytic feedback mechanism has been presented in Ref.~\cite{solomon_minsky_2013}.
The model is based on the Marshall-Walras equilibrium assumption for the price formation process.  It is able to properly reflect different economic regimes by the appropriate values of the parameters, which is also empirically validated in \cite{golo_too_2016}.
However, it model does not describe the transition process between different states of the economy,
which is the main point of interest in this paper. The model analyzed in \cite{solomon_minsky_2013} consists of several qualitatively similar but independent subparts that describe the dynamics of either loans or crisis accelerators. It does not factor in defaults and their influence on the interest rate during the loan accelerator phase, and, similarly, it does not factor in loans and their influence on the interest rate during the crisis accelerator phase. 

Within the model,  the dynamics of the loan number $N(t)$ and the interest rate value $i(t)$ is described as 
\begin{equation}
    \begin{aligned}
        N{\left(t \right)} &= \left(\frac{i{\left(t \right)}}{k}\right)^{- \mu}, \\
        i{\left(t + 1 \right)} &= i_0 N^{-\alpha} (t).
\end{aligned}
\label{eq:original_loans}
\end{equation}
Similarly, the dynamics of the crisis accelerator with the number of ponzi (i.e. technically defaulted) companies $D (t)$ and the interest rate $i (t)$ is defined as 
\begin{equation}
    \begin{aligned}
        D{\left(t \right)} &= \left(\frac{i{\left(t \right)}}{k}\right)^{\beta}, \\
        i{\left(t + 1 \right)} &= i_0 D^{\alpha} (t).
    \end{aligned}
    \label{eq:original_ponzi}
\end{equation}
A detailed justification of the equations is presented in \cite{solomon_minsky_2013}. Applied analysis includes probabilistic modeling of earnings and debt distribution in the market, a derivation from the power law assumption of wealth, earnings, and debt distribution, and a discussion of the power~law assumption. Empirical validation of the MWE model is performed in \cite{golo_too_2016}.
\begin{remark}
The model \cref{eq:original_loans} is one of the two loan dynamics models analyzed in \cite{solomon_minsky_2013}. It corresponds to the law of increasing returns.
Reasonability of the law of increasing returns in the banking industry is discussed in \cite{solomon_minsky_2013}. An additional reason for choosing it in this paper is the feature of a positive feedback loop: an increase in the number of loans causes a decrease in the interest rate, which in turn causes a further increase in the number of loans. Thus, the number of loans and interest rates are self-reinforcing. This feature of the system has interesting mathematical and real-world consequences, often poorly reflected in classical models of financial mathematics. 
\end{remark}


\subsection{Derivation of the unified model}
\label{subsec:model_derivation}

The aim is to unify the equations \cref{eq:original_loans}-\cref{eq:original_ponzi} into a single, consistent model (referred to as the unified MWE model, UMWE in short, throughout the rest of this work) that describes all variables of interest and their relations at each point in time. The proposed solution is as follows:
\begin{equation}
    \begin{aligned}
        N{\left(t \right)} &= \left(\frac{i{\left(t \right)}}{k}\right)^{- \mu}, \\
        D{\left(t \right)} &= \left(\frac{i{\left(t \right)}}{l}\right)^{\nu}, \\
        i{\left(t + 1 \right)} &= \frac{D^{\beta}{\left(t \right)}}{N^{\alpha}{\left(t \right)}},
    \end{aligned}
    \label{eq:my_model}
\end{equation}
where:
\begin{itemize}
    \item $N{\left(t \right)}$ is the volume of loans at time $t$;
    \item $D{\left(t \right)}$ -- the volume of defaults at time $t$;
    \item $i{\left(t + 1 \right)}$ -- the value of the interest rate at time $t$;
    \item $\mu$ is a parameter that describes the level of demand for credit in relation to the interest rate;
    \item $\nu$ --  the level of defaults in relation to the interest rate;
    \item $\alpha$ is a parameter that describes the sensitivity of the interest rate to the volume of loans;
    \item $\beta$ --  the sensitivity of the interest rate to the volume of defaults;
    \item $k, \ l$ are scale parameters. \newline
\end{itemize}

Define the set of utilised model parametrization as $\Lambda := \{ \alpha, \beta, \mu, \nu, k, l \} \in \mathbb{R}_{+}^{6}.$ As $k,l$ are technical scale parameters, they do not constitute a great point of interest in the investigation. Therefore, throughout the article, $\Lambda$ will be considered primarily as $ \Lambda =~\{ \alpha, \beta, \mu, \nu \} $ with the general exclusion of $k, l$ from the research focus.

The exponents in $\Lambda$ describe various economic quantities of the real world. The parameter $\mu$ reflects the dependence of the volume of loans on the interest rate and therefore the demand for credit for different levels of the interest rate. On the other hand, $\nu$ captures the relation between the volume of defaults and the interest rate, so it can be interpreted as the indicator of the resilience of a particular economy: systems with different values of the parameter $\nu$ are characterized by distinct default tendencies and fragility. The parameters $\alpha$ and $\beta$ describe the dynamics of the interest rate determination in response to the volumes of loans and defaults on the market. 

The dynamics of loans and defaults are taken directly from the original model. The interest rate is modeled in a natural way often met in finance as the (exponentiated) ratio of defaults to loans. Consider the case
\begin{equation}
    \alpha = \beta = 1,
\end{equation}
which implies that
\begin{equation}
    i(t+1) = \frac{D(t)}{N(t)}.
\end{equation}
This is a very simple and intuitive parameterization, where the interest rate equals the estimated probability of default $D_t / N_t.$ Moreover, the interest rate is also equal to the expected rate of return from the 'bare' credit with the same amount of loan and repayment. Consider a loan with the amount of $M$ and the same amount to repay at maturity $T.$ The expected rate of return from such a loan with the estimated probability of default $D_t / N_t$ and the recovery rate of $0$ equals
\begin{equation}
    r(t+1) = \frac{1}{M} \left( M \left( 1 - \frac{D(t)}{N(t)} \right) - M \right) = \left( 1 - \frac{D(t)}{N(t)} \right) - 1 = - \frac{D(t)}{N(t)}.
\end{equation}
Thus 
\begin{equation}
    i(t) = - r(t).
\end{equation}
The addition of $\alpha$ and $\beta$ parameters enhances the flexibility of the model and allows to reflect phenomena such as various market sentiments and premiums charged by banks for their offers. It also increases the accuracy potential to fit to real data. The proposed equation for the interest rate dynamics ultimately binds all variables of interest altogether.

It is worth noticing that if two of the power indices (either $\alpha$ and $\mu$ or $\beta$ and $\nu$) are set to $0,$ the result is effectively a two-equation model very similar to the original MWE. Furthermore, to accurately align both models, the initial interest rate component $i_0$ in \cref{eq:original_loans}-\cref{eq:original_ponzi} must satisfy $i_0=1.$ 
The exclusion of  $i_0$ from the interest rate equation ensures that the proposed model \cref{eq:my_model} is memoryless (Markovian). The initial interest rate $i_0$ was utilised in the original interest rate equations \cref{eq:original_loans}-\cref{eq:original_ponzi} as a scale parameter that anchors the evolution of the interest rate to its root. This causes $i_0$ to prevail in the model at each point in time and significantly influences the interest rate value even in the distant future. This long-memory approach has several drawbacks, including time inconsistency: the model parameterization and evolution are different depending on which point in time is treated as the initial moment. It is particularly problematic for the research conducted in this article, where one of the main points of interest is the transitions between qualitatively different economic regimes. It would be very inconvenient to handle the state transitions of the model with the long-lasting dependency on the predefined initial moment that prevails in every time step. The Markov property of the model \cref{eq:my_model} ensures that the value of the variable of interest in the next period depends only on the current one in the history, thus allowing for a very natural introduction of the regime switches.

\subsection{Mathematical analysis of the interest rate model}

Inserting defaults and loans into the interest rate definition in \cref{eq:my_model} yields a recurrence formula for $i(t):$
\begin{equation}
\label{eq:it_recurrence}
i\left(t+1 \right) =  \left( l^{- \beta \nu} k^{ - \mu \alpha } \right) \left( i{\left(t \right)} \right)^{ \alpha\mu + \beta \nu}.
\end{equation}
The fixed point $i_{fix}$ of the recurrence equation \cref{eq:it_recurrence} must satisfy
\begin{equation}
    i(t+1) = i(t) = i_{fix}. 
\end{equation}
Hence, it follows that
\begin{equation}
    \label{eq:i_fix}
    i_{fix} = \left( k^{- \alpha \mu} l^{- \beta \nu}\right)^{\frac{1}{1 - (\alpha \mu + \beta \nu)}}. 
\end{equation}
Just to recall, the fixed-point formula in the MWE model~\cite{solomon_minsky_2013} reads
\begin{equation}
    i^{MWE}_{fix} = \left( i_0 k^{- \alpha \mu} \right)^{\frac{1}{1 - \alpha \mu}}. 
\end{equation}
The fixed point in the unified model does not feature the dependency on the initial value of the interest rate $i_0.$ This is a noticeable enhancement of the model, as it is natural to expect that the invariant point of the system depends on the parameters values $\Lambda$ and should not be influenced by the initial value $i_0$ defined at an arbitrarily chosen point in time $t=0.$ In case of dependency of the fixed point on the initial rate $i_0,$ it is very hard to reasonably design transitions between different economic regimes, as the value of the invariant point of each new economy state is influenced by the arbitrarily chosen initial interest $i(0).$

Define the power index parameter 
\begin{equation}
    \label{eq:def_a}
    a :=  \mu \alpha + \nu \beta.
\end{equation}
From the recurrence equation \cref{eq:it_recurrence}, it follows that
\begin{equation}
\label{eq:it_recurrence_advanced}
    i \left(t+1 \right) =  \left( l^{- \beta \nu} k^{ - \mu \alpha } \right) i^{\left( \alpha\mu + \beta \nu \right)}{\left(t \right)}  = i_{fix}^{(1-a)}  i^{a}{\left(t \right)}.
\end{equation}
Formula \cref{eq:it_recurrence_advanced} is a simpler representation of the original interest rate recurrence equation. 
Consider now its logarithmic form.
\begin{equation}
\label{eq:it_recurrence_logarithmic}
    \ln (i(t+1)) = (1-a) \ln(i_{fix}) + a \ln(i(t)).
\end{equation}
Solving the last equation and returning to the nonlogarithmic form produces the compact formula  for the value of the interest rate $i(t)$ at any given point in time $t$, 
\begin{equation}
\label{eq:it_intime_t}
    i(t) = i_{fix} \left( \frac{i_{0}}{i_{fix}} \right)^{a^t}.
\end{equation}

\section{Credit market phases and transitions}
\label{sec:phase_transitions}

\subsection{Bifurcation analysis}
\label{subsec:bif_analysis}

The qualitative behavior of the system \cref{eq:my_model} depends mainly on two factors: the value of the power index $a$ and the position of the initial interest rate $ i (0) $ relative to the fixed point $ i_{fix}$. The conclusions presented can be straightforwardly derived from the formula for the interest rate over time \cref{eq:it_intime_t}. For a general overview of dynamic systems, fixed points, and their classification, see \cite{strogatz_nonlinear_2007}.

\begin{figure}[htbp]
  \centering
  \label{fig:stable}
  \includegraphics[scale=0.5]{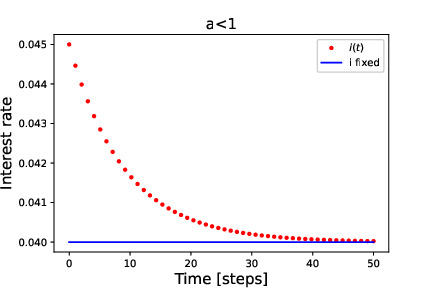}
  \caption{Stable regime with $ a = 0.9, \ i_0 = 0.045, \ i_{fix} = 0.04.$}
  \label{fig:c_stable}
\end{figure}

For $ a < 1$ in the Eq.~\cref{eq:it_intime_t} the fixed point is attractive (\cref{fig:c_stable}). This situation corresponds to a stable and healthy economic regime with interest rates that smoothly and monotonically converge to the fixed point $ i_{fix}. $ From the perspective of central bank seeking for stability, this is the regime that should be searched for. 

The point $ a = 1 $ is the point of transcritical bifurcation of the system. The behavior of the system changes drastically around this point.

For $a=1$, from the recurrence equation \cref{eq:it_recurrence} it follows that
\begin{equation}
    \begin{aligned}
        i\left(t+1 \right) &= \left( l^{- \beta \nu} k^{ - \mu \alpha } \right)  i^a{\left(t \right)} \\
        &= \left( l^{- \beta \nu} k^{ - \mu \alpha } \right) i{\left(t \right)}  \\
        &= \left( l^{- \beta \nu} k^{ - \mu \alpha } \right)^2 i{\left(t-1\right)} \\
        &= \left( l^{- \beta \nu} k^{ - \mu \alpha } \right)^n  i{\left(t-n + 1 \right)} \\ 
        &= \left( l^{- \beta \nu} k^{ - \mu \alpha } \right)^{t+1} i_{0} \\
        &= c^{t+1} i_{0}, \qquad c := \left( l^{- \beta \nu} k^{ - \mu \alpha } \right). \\
    \end{aligned}
    \label{eq:it_a=1}
\end{equation}
If $ c = 1$, then 
\begin{equation}
    i\left(t \right) =  \left( l^{- \beta \nu} k^{ - \mu \alpha } \right)  i{\left(t \right)}  = c i{\left(t \right)} = i{\left(t \right)}.
\end{equation}
Therefore, the interest rate is constant over time and equals the initial value:
\begin{equation}
    i\left(t \right) = i_0.
\end{equation}
For $ c \neq 1,$ there is no fixed point of the equation \cref{eq:it_a=1} (with the exception of $0,$ which is generally not considered a possible value of the interest rate in the analyzed model). If $c<1,$ interest rate converges to $ 0 $ and if $c>1,$ interest rate diverges to infinity.

For $ a > 1, $ the fixed point is repelling. This situation corresponds to a dangerous and unstable state of the economy, with the interest rate rapidly approaching $0$ or infinity. Contrary to the stable regime $ a < 1 $, the behavior of the system is heavily dependent on the position of the initial interest rate $i_0$ relative to the fixed point $i_{fix}.$ 
For $ i_{0} > i_{fix}, $ it follows that
\begin{equation}
    \begin{aligned}
        i_0 &> \left( k^{- \alpha \mu} l^{- \beta \nu}\right)^{\frac{1}{1 - a}}  \\
        \ln i_0 &> ( - \alpha \mu \ln k  - \beta \nu \ln l)/(1-a) \\
        0 &> \frac{1}{1-a} ( - \alpha \mu \ln k  - \beta \nu \ln l  - (1-a) \ln i_0) \\
        0 &> \frac{1}{1-a} \left(  \alpha \mu ( \ln i_0 - \ln k )  + \beta \nu (\ln i_0 - \ln l ) - \ln i_0 \right). \\
    \end{aligned}
\end{equation}
The last form of the above formula will be referred to as the position inequality,
\begin{equation}
    \label{ineq:position}
    0 > \frac{1}{1-a} \left(  \alpha \mu ( \ln i_0 - \ln k )  + \beta \nu (\ln i_0 - \ln l ) - \ln i_0 \right),
\end{equation}
which determines the qualitative behavior of the system in the unstable regime $a>1.$

\begin{figure}[htbp]
  \centering
  \label{fig:bubble}
  \includegraphics[scale=0.5]{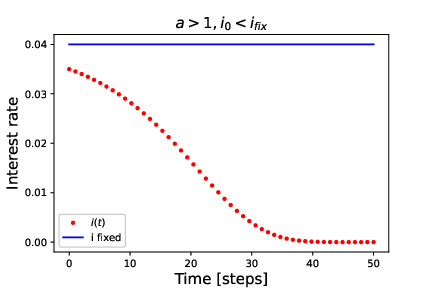}
  \caption{Bubble regime with $ a = 1.1, \ i_0 = 0.035, \ i_{fix} = 0.04.$}
  \label{fig:c_bubble}
\end{figure}

If $ i_{0} < i_{fix}, $ $ i(t) $ converges to $0$ (\cref{fig:c_bubble}). We say that the economy is in a bubble regime. This is an unsafe and unsustainable state. According to the model \cref{eq:my_model}, the volume of loans $N(t),$ and therefore the amount of money in the economy, is inversely proportional to the interest rate: $N{\left(t \right)} =  \left( k / i{\left(t \right)} \right)^{\mu}. $ As interest rate approaches $0,$ the amount of money in the systems rapidly expands without constraints. Inflation spirals out of control and either the system collapses or the currency fails to fulfill a basic function of money: a stable measure of value \cite{stanley_jevons_money_1989}. 

\begin{figure}[htbp]
  \centering
  \label{fig:crash}
  \includegraphics[scale=0.5]{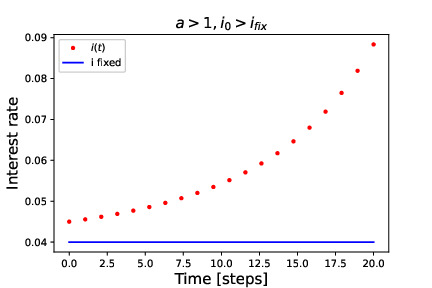}
  \caption{Crash regime with $ a = 1.1, \ i_0 = 0.045, \ i_{fix} = 0.04.$}
  \label{fig:c_crash}
\end{figure}

If $ i_{0} > i_{fix}, $ $ i(t) $ diverges to infinity (\cref{fig:c_crash}).
The economy is in a crash regime. The interest rate increases sharply, entering the feedback loop with vanishing credit and expanding defaults. The collapse of the banking system has very serious consequences beyond the limitation of the credit activity. In the modern economy, a large part of the money itself is a credit created by banks \cite{mcleay_money_2014}. Therefore, the collapse of the banking system is the collapse of money itself. 

Overall, the UMWE model has features similar to those of the original model. However, it extends the results of the original one to a holistic framework, describing the comprehensive dynamics of the credit market with loans, defaults, and interest rates at each point in time. On the other hand, the credit cycle is a sum of its phases with the transition between them. The UMWE model is able to explain the stable growth, bubble, and crisis phases, depending on the values of the parameters. Therefore, to explain a credit cycle, the model has to allow for changes in the parameters in such a way that the transition occurs between different states of the system.

\subsection{Critical parameters}
\label{sec:bifurcation_paramaters_analysis}

In the dynamic parameter regime, there is a time dependency of the parameter values $\Lambda (t) := \left \{ \alpha(t), \beta(t), \mu(t), \nu(t) \right \}. $ However, the unified MWE model \cref{eq:my_model} is Markovian. Therefore, for the sake of clarity, the time dependence of the parameters will be omitted from the notation. The value of the parameter can be interpreted as being at a "current" or any arbitrary chosen point in time.

Recall that  $a=\alpha \mu + \beta \nu .$  Transcritical bifurcation occurs at $a=1.$ Therefore, the critical value $\alpha_{crit}$ of $\alpha,$ for which the bifurcation occurs must satisfy
\begin{equation}
\label{eq:critical_alpha_condition}
    \alpha_{crit} \mu + \beta \nu = 1.
\end{equation}
Thus, the value of the critical stability parameter $\alpha_{crit}$ equals 
\begin{equation}
    \label{eq:critical_alpha}
    \alpha_{crit} = \frac{1 - \beta \nu}{\mu}.
\end{equation}
The application of the same reasoning to the other parameters yields the following values of the critical stability parameters:
\begin{equation}
    \begin{aligned}
        \alpha_{crit} &= \frac{1 - \beta \nu}{\mu}, \\
        \beta_{crit}  &=   \frac{1 - \alpha \mu}{\nu}, \\
        \mu_{crit} &= \frac{1 - \beta \nu}{\alpha}, \\
        \nu_{crit}  &=  \frac{1 - \alpha \mu}{\beta}. \\
    \end{aligned}
    \label{eq:crtiical_stability_parameters}
\end{equation}

At an aggregated level, the value $\Delta_{crit}$ of a scale parameter can be defined, for which a stability transition occurs. Let $ \Delta_{crit}$ be a multiplier for which a critical value $\alpha \mu + \beta \nu = 1$ is attained:
\begin{equation}
    \Delta_{crit}  \alpha \Delta_{crit}  \mu   + \Delta_{crit} \beta \Delta_{crit} \nu = 1.
\end{equation}
Hence, it follows that $\Delta_{crit}$ satisfies
\begin{equation}
    \begin{aligned}
    \Delta^{2}_{crit} \left(\alpha \mu + \beta \nu \right) &= 1 \\
    \Delta_{crit} &= \frac{1}{\sqrt{a}}.
    \end{aligned}
    \label{eq:critical_delta}
\end{equation}

On the other hand, consider the transition between bubble and crash regimes in the unstable state $a>1.$ The regime is determined by the relative positions of the current interest rate $i_t$ and the fixed point $i_{fix}.$ Therefore, the value of the critical direction parameter $\alpha_{crit} (i_t)$ for which there is a change of the sign in parentheses term in the position inequality \cref{ineq:position} equals
\begin{equation}
    \label{eq:direction_alpha}
    \alpha_{crit} (i_t) =  \frac{\ln i_t -  \beta \nu ( \ln i_t - \ln l )}{\mu (\ln i_t - \ln k )}.
\end{equation}
In the unstable regime $a>1$, either $ i_t \longrightarrow 0$ or $ i_t \longrightarrow \infty $ as $ t \longrightarrow \infty. $ Hence, the limit $\overline{\alpha_{crit}}$ of the above formula \cref{eq:direction_alpha} as $t$ approaches infinity is
\begin{equation}
    \overline{\alpha}_{crit} = \frac{1 - \beta \nu}{\mu}.
\end{equation}
It is worth noting that it is the same value as the critical stability parameter $\alpha_{crit}$ in Eq.~\cref{eq:critical_alpha}.

The critical direction values of the other parameters can be derived analogously, yielding
\begin{equation}
    \begin{aligned}
        \alpha_{crit} (i_t) &=  \frac{\ln i_t -  \beta \nu ( \ln i_t - \ln l )}{\mu (\ln i_t - \ln k )}, \\
        \beta_{crit} (i_t) &=  \frac{\ln i_t -  \alpha \mu ( \ln i_t - \ln k )}{\nu (\ln i_t - \ln l )}, \\
        \mu_{crit} (i_t) &=  \frac{\ln i_t -  \beta \nu ( \ln i_t - \ln k )}{\alpha (\ln i_t - \ln l )}, \\
        \nu_{crit} (i_t) &=  \frac{\ln i_t -  \alpha \mu ( \ln i_t - \ln k )}{\beta (\ln i_t - \ln l )}. \\ 
    \end{aligned}
    \label{eq:critical_direction_parameters}
\end{equation}
In the limit cases (either $ i_t \longrightarrow 0$ or $ i_t \longrightarrow \infty $), the asymptotic critical direction parameters equal
\begin{equation}
    \begin{aligned}
        \overline{\alpha}_{crit} &= \frac{1 - \beta \nu}{\mu}, \\
        \overline{\beta}_{crit}  &=   \frac{1 - \alpha \mu}{\nu}, \\
        \overline{\mu}_{crit} &= \frac{1 - \beta \nu}{\alpha}, \\
        \overline{\nu}_{crit}  &=  \frac{1 - \alpha \mu}{\beta}. \\
    \end{aligned}
\end{equation}
They are exactly the same as the values of critical stability parameters in Eq.~\cref{eq:crtiical_stability_parameters}.

Furthermore, at a general level, define the value $\Delta_{crit} (i_t)$ of a scale parameter for which a direction transition occurs. Let $ \Delta_{crit} (i_t)$ be such that in the position inequality \cref{ineq:position} $0$ is attained:
\begin{equation}
    \begin{aligned}
        0 &= \Delta_{crit} (i_t) \alpha \Delta_{crit} (i_t)  \mu ( \ln i_t - \ln k ) + \\ 
        &+ \Delta_{crit} (i_t) \beta \Delta_{crit} (i_t) \nu (\ln i_t - \ln l ) - \ln i_t.
    \end{aligned}
\end{equation}
Therefore
\begin{equation}
    \begin{aligned}
    \label{eq:direction_delta}
        \Delta^{2}_{crit} (i_t) &=  \frac{\ln i_t}{\alpha \mu ( \ln i_t - \ln k )  + \beta  \nu (\ln i_t - \ln l )} \\
        \Delta_{crit} (i_t) &=  \sqrt{ \frac{\ln i_t}{\alpha \mu ( \ln \frac{i_t}{k} )  + \beta  \nu (\ln \frac{i_t}{l} )}}.
    \end{aligned}
\end{equation}
In the limit case $ t \longrightarrow \infty $, the value of the asymptotic critical direction delta $\overline{\Delta}_{crit}$ is equal to the critical stability delta $\Delta_{crit}$ in Eq.~\cref{eq:critical_delta}:
\begin{equation}
    \label{eq:direction_delta_limit}
    \overline{\Delta}_{crit} =  \sqrt{ \frac{1}{\alpha \mu  + \beta  \nu }} = \frac{1}{\sqrt{a}}.
\end{equation}

\begin{remark}
Similar definitions can be introduced in the stable regime $a<1$, but the position of $i_{fix}$ relative to $i_t$ is not so relevant there. In that regime, $i_t$ converges to $i_{fix}$ in all cases, so the qualitative behavior of the stable system does not change significantly depending on the location of the current interest rate (with the exception of the direction of convergence).
\end{remark}

The comparison of critical stability parameters to asymptotic critical direction parameters performed in the analysis makes it possible to deduce the following result.
\begin{theorem}[Equivalence of the Critical Parameters]
\label{th:critical_parameters_equivalence}
Let $\Lambda = \left \{ \alpha, \beta, \mu, \nu \right \}$ be the set of exponential parameters of the UMWE model \cref{eq:my_model}. 
For every parameter $\lambda \in \Lambda$ and every pair of the corresponding critical stability and asymptotic critical direction parameters $\left( \lambda_{crit}, \overline{\lambda}_{crit} \right)$ the following relation takes place:
\begin{equation}
    \lambda_{crit} = \overline{\lambda}_{crit}.
\end{equation}
Furthermore, it holds that 
\begin{equation}
    \Delta_{crit} = \overline{\Delta}_{crit}.
\end{equation}
\end{theorem}
The result of \cref{th:critical_parameters_equivalence} is not a coincidence. Consider the below form of the position inequality \cref{ineq:position}:
\begin{equation}
     0 > \frac{1}{1-a} ( - \alpha \mu \ln k  - \beta \nu \ln l  - (1-a) \ln i_t).
\end{equation}
In the unstable regime, $ \ln i_t $ approaches $+ \infty$ or $- \infty$. Therefore, in the limit case, the parameter $ - \alpha \mu \ln k  - \beta \nu \ln l $ is negligible; to change the sign in the formula, the sign of $(1-a)$ should be changed. 

The result of the \cref{th:critical_parameters_equivalence} unifies asymptotic critical direction parameters with critical stability parameters, highlighting the latter as the most important indicators of the fragility of the analyzed UMWE model. Thus, the neighborhood of the point $a=1$ constitutes indeed the critical area of the system, where the phase transitions (between stability and instability, as well as between a bubble and a crisis) ultimately occur. 

\subsection{Measures of systemic risk}

On the basis of the determined critical (stability and direction) parameters, the measures of systemic risk for the unified MWE model can be constructed as the distances of the current values of the parameters to the critical ones. 
For every parameter $ \lambda \in \Lambda $ and the corresponding critical stability parameter $\lambda_{crit}$ define 

\begin{equation}
    \label{eq:stability_distances}
    \begin{aligned}
        \delta \lambda_{crit} &= \lambda_{crit} - \lambda, \\
        \frac{\delta \lambda_{crit}}{\lambda} &= \frac{\lambda_{crit} - \lambda}{\lambda} = \frac{\lambda_{crit}}{\lambda} - 1.
    \end{aligned}
\end{equation}
The exact formulas for all parameters are presented in \cref{tab:critical_stability_distances}. The above measures will be referred to as the (critical) stability distances: absolute stability distance $\delta \lambda_{crit}$ and relative one, $\frac{\delta \lambda_{crit}}{\lambda}.$ According to the theorem \cref{th:critical_parameters_equivalence}, distances to critical stability parameters are equal to the distances to asymptotic critical direction parameters.
\begin{table}[h!]
\label{tab:critical_stability_distances}
\centering
 \begin{tabular}{|c | c c c c c|} 
 \hline
& $\alpha$ & $\beta$ & $\mu$ & $\nu$ & $\Delta$ \\ 
 \hline
 $\delta \lambda_{crit}$ & $\frac{1-a}{\mu}$ & $ \frac{1-a}{\nu} $ & $\frac{1-a}{\alpha}$ & $\frac{1-a}{\beta}$  & $\frac{1}{\sqrt{a}} - 1$ \\ [1ex] 
$\delta \lambda_{crit} / \lambda$& $ \frac{1-a}{\mu \alpha}$ & $\frac{1-a}{\nu \beta}$ & $\frac{1-a}{\mu \alpha}$ & $\frac{1-a}{\nu \beta}$  & $\frac{1}{\sqrt{a}} - 1$ \\ [1ex] 
 \hline
 \end{tabular}
 \caption{\label{demo-table}Critical stability distances.}
\end{table}
For critical direction parameters, the distances can be defined analogously for $ \lambda \in \Lambda $:
\begin{equation}
    \label{eq:direction_distances}
    \begin{aligned}
        \delta \lambda_{crit} (i_t) &= \lambda_{crit} (i_t) - \lambda, \\
        \frac{\delta \lambda_{crit} (i_t)}{\lambda} &= \frac{\lambda_{crit} (i_t) - \lambda}{\lambda} = \frac{\lambda_{crit} (i_t)}{\lambda} - 1.
    \end{aligned}
\end{equation}
The exact formulas for all parameters are presented in \cref{tab:critical_direction_distances}. In that case, $\delta \lambda_{crit} (i_t)$ and $\frac{\delta \lambda_{crit} (i_t)}{\lambda}$ are referred to as absolute and relative (critical) distances, respectively. 
\begin{table}[h!]
\label{tab:critical_direction_distances}
\centering
 \begin{tabular}{|c | c c c|} 
 \hline
& $\alpha$ & $\beta$ & $\mu$ \\ 
 \hline
 $\delta \lambda_{crit} \left( i_t \right)$ & $- \frac{\beta \nu \ln \frac{i_0}{l} - \ln i_0}{\mu \ln \frac{i_0}{k}} - \alpha $ & $- \frac{\alpha \mu \ln \frac{i_0}{l} - \ln i_0}{\nu \ln \frac{i_0}{k}} - \beta $ & $ - \frac{\beta \nu \ln \frac{i_0}{l} - \ln i_0}{\alpha \ln \frac{i_0}{k}} - \mu$ \\ 

$\delta \lambda_{crit} \left( i_t \right) / \lambda$ & $- \frac{\beta \nu \ln \frac{i_0}{l} - \ln i_0}{\alpha \mu \ln \frac{i_0}{k}} - 1 $ & $- \frac{\alpha \mu \ln \frac{i_0}{l} - \ln i_0}{ \beta \nu \ln \frac{i_0}{k}} - 1$ & $- \frac{\alpha \mu \ln \frac{i_0}{l} - \ln i_0}{ \beta \nu \ln \frac{i_0}{k}} - 1$ \\ 
\hline
& $\nu$ & $\Delta$ &   \\
\hline
$\delta \lambda_{crit} \left( i_t \right)$ & $ - \frac{\alpha \mu \ln \frac{i_0}{l} - \ln i_0}{\beta \ln \frac{i_0}{k}} - \nu$ & $\sqrt{ \frac{\ln i_t}{\alpha \mu ( \ln \frac{i_t}{k} )  + \beta  \nu (\ln \frac{i_t}{l} )}}$ &   \\
$\delta \lambda_{crit} \left( i_t \right) / \lambda$ & $ - \frac{\alpha \mu \ln \frac{i_0}{l} - \ln i_0}{ \beta \nu \ln \frac{i_0}{k}} - 1 $ & $\sqrt{ \frac{\ln i_t}{\alpha \mu ( \ln \frac{i_t}{k} )  + \beta  \nu (\ln \frac{i_t}{l} )}}$ &  \\
\hline
 \end{tabular}
 \caption{\label{table} Critical direction distances.}
\end{table}

System fragility measures defined in \cref{eq:stability_distances} and \cref{eq:direction_distances} are of very practical use. Stability distances describe how far (from the perspective of values of parameters) the system is from the unstable (or stable) state. In a stable stage, this information can be utilised to keep a (arbitrary set) safe distance from the unstable phase. But, as usually in finance  there could be a trade-off between the risk and the return \cite{markowitz_portfolio_1952}: lower values of the parameters can mean a higher level of interest rates $i_{fix}.$ For unstable phase, the stability distance can be utilized to steer the comeback to the stable regime. 

On the other hand, direction distance describes how far the system is from the crisis (or bubble) in the unstable regime $ a > 1.$ It can be helpful to determine how to escape the bubble with too much money present in the economy, causing inflation, instability, and phantom (nominal) growth without coverage in real terms. It is worth to point out that it is not obvious if escaping bubble smoothly in the long period is a better solution than v-shaped crisis and recovery. Smooth bubble escape could mean long lasting, exhausting recession with prevailing consequences, often difficult to get rid off. In comparison, a rapid, severe, but quick shock, controlled to some extent by central bank supporting the financial situation, can often be followed by relatively fast recovery to the stable regime. With enough knowledge regarding the fit of the presented model to reality, possibly the direction distance could be utilized to steer a more rapid recovery and then smooth transition to the stable regime. For these purposes, however, empirical validation of the model is required.

\section{The credit cycle}
\label{sec:cycle}

\subsection{Theoretical background}

\begin{figure}
    \centering
    \includegraphics[scale=0.5]{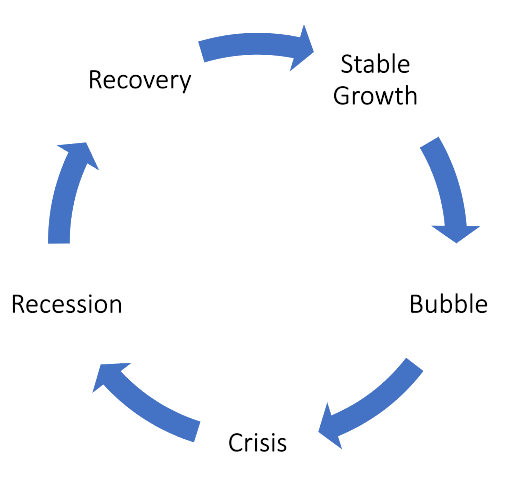}
    \caption{An example credit cycle with various phases. }
    \label{fig:graph_credit_cycle}
\end{figure}

The credit cycle is the expansion and reduction of credit in the economy over time \cite{phillips_banking_1938}. It may consist of various stages such as expansion, crisis, recession, and recovery (\cref{fig:graph_credit_cycle}).
 Excessive growth of the money supply and the level of prices can lead to the formation of a speculative bubble, which is unsustainable in the long term and may result in a severe crash \cite{benning_trading_2007}. The crisis is usually followed by a period of decline in credit and economic activity, which eventually stabilizes the situation and leads to the next phase of expansion. Throughout the cycle, different states are characterized by very different values of economic and financial factors, such as interest rates, inflation, unemployment, industrial production, money supply and bankruptcies \cite{wankel_encyclopedia_2009}. This diversity is captured by the UMWE model \cref{eq:my_model}, heavily dependent on parameterization $\Lambda,$ where for various sets of parameters the system can exhibit  qualitatively different types of dynamics, as analyzed in \cref{subsec:bif_analysis}.

The parameters $\alpha$ and $\beta$  reflect the policy of the banks regarding their market offer, expressed through the price charged for their products. Since this article is written mainly from the point of view of financial institutions, it will be assumed that $\alpha$ and $\beta$ could change their values throughout the credit cycle, as they represent the available instruments of the financial policy and can be tailored to individual situations. The values of the parameters $\mu$ and $\nu$ will be treated as external from the bank's perspective and therefore kept constant. The technical scale parameters $k, \ l$ will also not vary over time in the proposed approach. The assumption that only $\alpha$ and $\beta$ can change over time simplifies the analysis and makes it more clear. Nevertheless, it is worth noting that during the incorporation of the model into the real data analysis, the potential dynamics of the rest of the parameters should also be validated, possibly even with the relations between various parameters, as for example $\alpha$ and $\mu$ could both be driven by the common optimism in the credit market.

\subsection{Stable phase and the bubble}
\label{sec:stable_and_bubble}

It is argued in Ref. \cite{solomon_minsky_2013} that the square root power law can describe the response of the economic demand for loans in relation to the interest rate. Based on that, the values of exponents that describe the dependence of loans and defaults on the interest rates in the system \cref{eq:my_model} are set to 
\begin{equation}
    \mu=\nu=0.499.
\end{equation}
The values are just below $0.5$ for technical reasons explained later.

For the exponent parameters of the interest rate equation in the system \cref{eq:my_model}, the initial values are chosen as
\begin{equation}
    \alpha = \beta = 1,
\end{equation}
which implies that the interest rate equals the estimated probability of default
\begin{equation}
    i_{t+1} = \frac{D_t}{N_t}.
\end{equation}
The rationale for such a parameterization was discussed in \cref{subsec:model_derivation}.
The scale parameters are set to 
\begin{equation}
    \begin{aligned}
        k &= 105.5, \\
        l &= 0.0096.
    \end{aligned}
\end{equation}
They are just technically adjusted to fit the desired behavior of the model. Finally, the initial value of the interest rate is taken as
\begin{equation}
    i_0 = 0.42.
\end{equation}

\begin{figure}[htbp]
  \centering
  \includegraphics[scale=0.5]{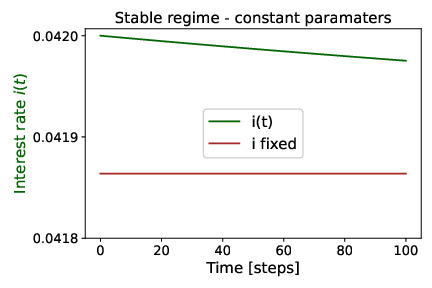}
    \caption{Stable phase of the cycle with $ \alpha = \beta = 1,  \ \mu = \nu = 0.499, \ k = 105.5, l = 0.0096, \ i_0 = 0.042, \ i_{fix} \approx  0.04186.$ The evolution of the system starts at a stable and calm phase. The interest rate slowly and smoothly settles to its lower limit.}
  \label{fig:cycle_stable}
\end{figure}

The chosen specification of the parameters implies that
\begin{equation}
    a = \alpha \mu + \beta \nu = 1 \times 0.499 + 1 \times 0.499 = 0.998 < 1.
\end{equation}
It means that the default state of economy is a convergent one ($a<1$), but it is very close to the instability border $a=1$ at the same time. The value of the fixed interest rate of such parameterized system is close to $0.04186.$ Therefore, the evolution of the system starts at a stable and calm phase. The interest rate slowly and smoothly settles to its lower limit.
The described situation is illustrated in \cref{fig:cycle_stable}.

\begin{figure}[htbp]
  \centering
   \begin{subfigure}[b]{0.55\textwidth}
  \centering
  \includegraphics[align=c,width=\textwidth]{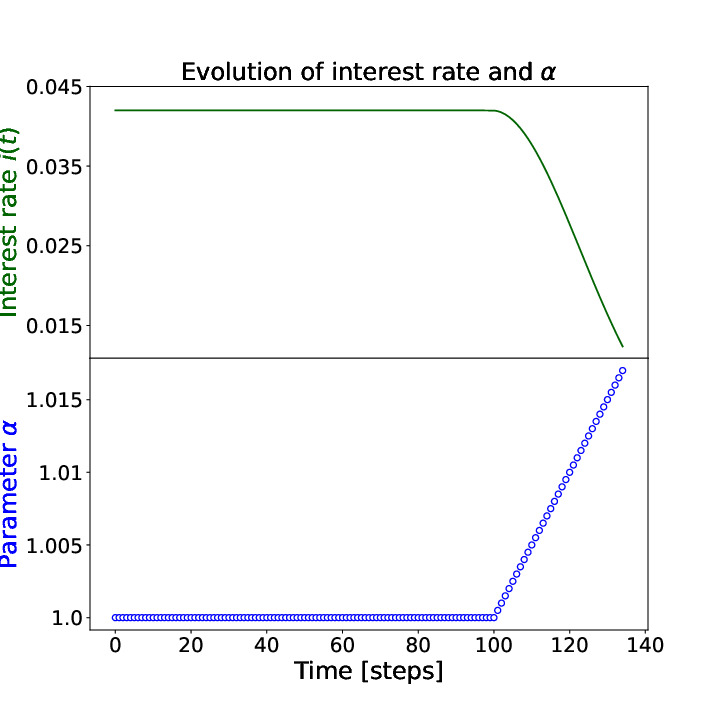}
  \end{subfigure}
  \hfill
  \begin{subfigure}[b]{0.43\textwidth}
  \centering
  \includegraphics[align=c,width=\textwidth]{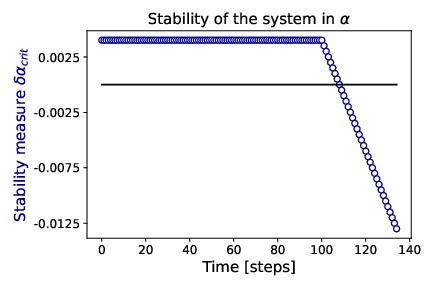}
  \end{subfigure}
  \caption{Bubble phase of the cycle with $  \beta = 1,  \ \mu = \nu = 0.499, \ k = 105.5, l = 0.0096$ and increasing $\alpha.$ Long period of stable growth improved market confidence, expressed in rising value of $\alpha$ (left plot). System stability decreases and at some point in time market enters the bubble phase (right plot).}
  \label{fig:cycle_bubble}
\end{figure}

The prevailing period of prosperity amplifies the confidence and optimism of the banks. At some point in time, they become comfortable enough to start to relax the previous rules of determining the value of the interest rate in relation to loans and defaults. In the selected setup of the parameters $\Lambda,$ the interest rates could be lowered by the increase of $\alpha$ or the decrease of $\beta.$ Modification of the sensitivity to the number of loans (money) in the system appears to be less risky than modification of the sensitivity to the number of defaults. Therefore, after a long period of stable economic growth, $\alpha$ increases. Our approach brings positive feedback from the economy, resulting~in more beneficial dynamics of the credit market in comparison to the previous one. The favorable outcome fuels further optimism and encourages to increase $\alpha$ even more. The result is a positive feedback loop with the parameter $\alpha$ increasing at each step. Along with the rise of the value of $\alpha$, the distance to instability $\delta \alpha_{crit}$ decreases and the system becomes more fragile. At some point in time, $\delta \alpha_{crit}$ passes through $0.$ The market enters a bubble state and continues to grow. The situation is presented in \cref{fig:cycle_bubble}.

Although it is generally assumed that lower interest rates fuel economic growth \cite{keynes_general_2018}, too low levels of interest rates are not healthy for the economy. Interest rates influence the supply of money \cite{ krugman_macroeconomics_2009, mankiw_principles_2009,mcleay_money_2014}, which is also reflected in the considered model. Interest rates impact the volume of loans and, in turn, the amount of money in the economy, by the relation given in \cref{eq:my_model}:
\begin{equation}
    N{\left(t \right)} = \left(\frac{i{\left(t \right)}}{k}\right)^{- \mu}.
\end{equation}
Furthermore, the increase in the amount of money leads to the increase in the level of prices \cite{palgrave_macmillan_equation_2008} (there are also direct relations between interest rates and inflation derived in economic theory \cite{fisher_nature_2009}). Exceeding the unhealthy threshold of the inflation level creates the risk of entering the vicious cycle of self-reinforcing increases in prices and wages \cite{mankiw_brief_2008}. Rising prices increase the value of collateral for credit, improving the credit quality of borrowers and amplifying the increase in credit volume in the economy \cite{bernanke_financial_1996}. This can create incentives for the formation of speculative bubbles, which amplifies the fragility of the financial system \cite{kiyotaki_credit_1997} and potentially leads to a crash, as was the case in 2007 \cite{krishnamurthy_amplification_2010}. When the interest rate reaches near zero territories in the bubble regime, the amount of money in the system and the level of prices expand uncontrollably without limitations; there are known cases of yearly inflation reaching levels of $10^{22} \%$ \cite{hanke_measurement_nodate}. The currency no longer performs the function of a measure and a store of value, violating the definition of money itself \cite{stanley_jevons_money_1989}.

\subsection{Crisis and stabilization}

As discussed in \cref{sec:stable_and_bubble}, there are practical limits on growth of the monetary base, and therefore the interest rate should not remain in the near-zero territory indefinitely (there are some exceptions even with negative interest rates \cite{altavilla_is_2022, blanchard_public_2019,blinder_revisiting_2012,heider_life_2019}, but they are beyond the scope of this work). Central banks are generally obliged to preserve the stable value of the currency and therefore maintain inflation at predefined acceptable levels \cite{levy_yeyati_monetary_2010}. Steering interest rate levels remains a key tool of central bank monetary policy implementation \cite{chenery_handbook_1988, smelser_international_2001}, therefore, interest rates are risen to bring inflation back to target levels. The lower bound on the interest rate offered by banks in the market is the rate of return on their deposits placed in the central bank accounts. This value, called the deposit rate, is established by the central bank \cite{bernanke_new_2020}. Other limitations come from the Basel accords, placing capital and liquidity requirements, as well as the permissible leverage ratio \cite{grundke_impact_2020, basel_comittee_on_banking_supervision_basel_nodate}.

\begin{figure}[htbp]
  \centering
  \begin{subfigure}[b]{0.43\textwidth}
  \centering
  \includegraphics[width=\textwidth]{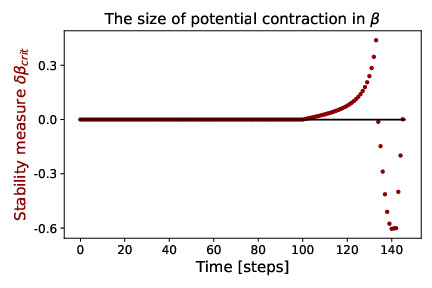}
  \includegraphics[width=\textwidth]{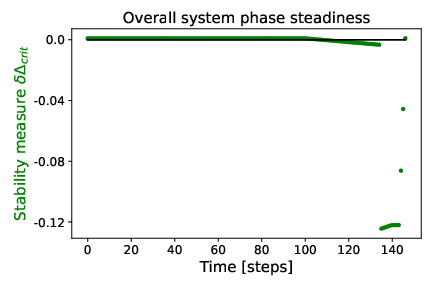}
  \end{subfigure}
  \hfill
  \begin{subfigure}[b]{0.55\textwidth}
  \centering
  \includegraphics[width=\textwidth]{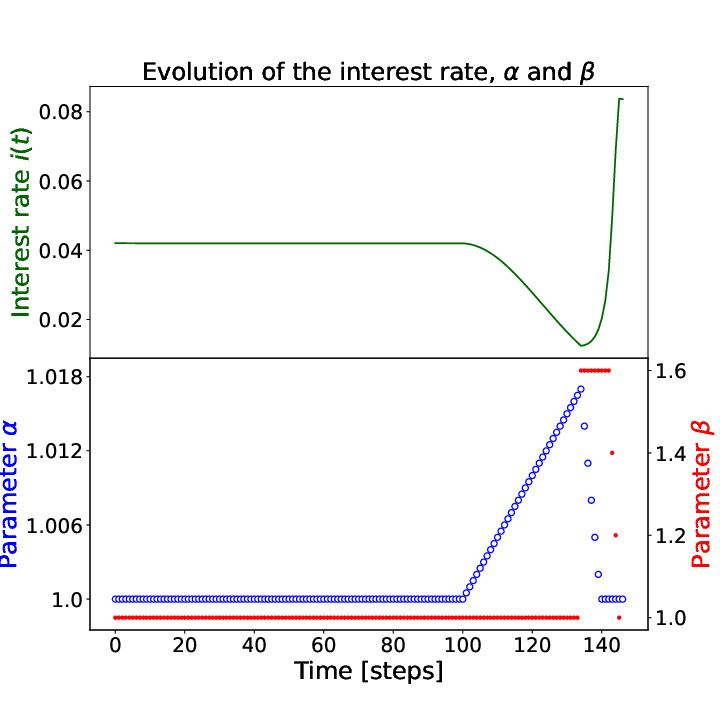}
  \end{subfigure}
  \hfill
  \caption{Crisis phase of the cycle with $  \beta = 1.6,  \ \mu = \nu = 0.499, \ k = 105.5, l = 0.0096$ and decreasing $\alpha.$ Left column: the top chart presents the severity of potential contraction of the system expressed in terms of value of critical direction distance for $\beta$. The bottom chart presents the overall system phase steadiness expressed in terms of critical direction distance $\delta \Delta_{crit}$. Right column: The values of $\alpha$ are put on the left vertical axis of the bottom plot, whereas the ones of $\beta$ - on the right axis. The system comes close to the practical minimum value of interest rate, accumulating excessive financial material for the collapse. The result is the shock of $\beta$ parameter and reverse in the direction of interest rate evolution. The overall system phase steadiness changes dynamically over time.}
  \label{fig:cycle_crash}
\end{figure}

All this leads to the conclusion that there are practical restrictions on the decrease of the interest rate, reflected in the adopted approach by the minimum value for which banks will not allow the interest rate to fall below, $0.123.$
The situation discussed here is presented in \cref{fig:cycle_crash}.
When the interest rate approaches the limit, banks start to fear. In order to remain safe, they increase the sensitivity  to the number of defaults, $\beta$, in order to distance the interest rate from the limit level. Because $\beta$ increases and $ a = \alpha \mu + \beta \nu,$ the market remains in the unstable regime $a>1$. But to avoid danger, banks rise $\beta$ to such a level that the distance from the minimum interest rate increases as well.  This increase in conjunction with the presence of the unstable regime $a>1$ implies a crash.

For the transition between bubble and crash to occur, the critical direction distance  
\begin{equation}
    \delta \beta_{crit} =  \frac{\ln i_t -  \alpha \mu ( \ln i_t - \ln k )}{\nu (\ln i_t - \ln l )} - \beta
\end{equation}
must fall below $0$.
The risk measure $\delta \beta_{crit}$ is a linear function of $\alpha$ with the slope of $  \frac{ -   \mu ( \ln i_t - \ln k )}{\nu (\ln i_t - \ln l )} $
and a hyperbolic function of the current value of the interest rate, $ i_t. $ It is qualitatively dependent on values of $k, \ l$ parameters, as they determine the signs of appropriate fragments of the equation above and then the monotonicity of dependencies between the exponents of the model.
This implied from the model formula, in the set-up of the chosen parameters, reflects an important feature of reality: the faster the bubble grows (greater $\alpha$), the faster and severe the contraction, because $a = \alpha \mu + \beta \nu$ describes the velocity of divergence.

The transition to the crash regime results in a sharp increase in the volume of defaults and a significant contraction in the volume of loans. The crunch of the loan market amplifies the panic, causing banks to further reduce lending by decreasing the value of $\alpha,$ which ultimately falls to its initial level. The distance to stability $\Delta_{crit}$ starts to decrease in absolute terms, consequently driving the system closer to the stable state. Meanwhile, the central bank has time to intervene and calm down the situation. After bringing back the sensitivity $\alpha$ to territories known from the stable regime, banks start to reduce the shocked parameter $\beta.$ Finally, $\Delta_{crit}$ passes through $0,$ parameters return to their initial level and the system returns to the stable state.

\begin{figure}[htbp]
  \centering
  \includegraphics[scale=0.3]{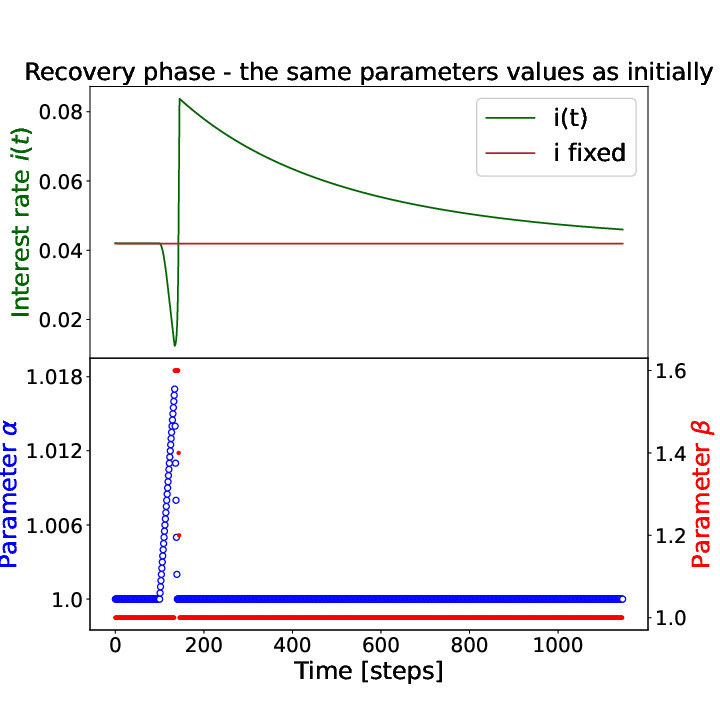}
  \caption{Recovery phase of the cycle with $ \alpha = \beta = 1,  \ \mu = \nu = 0.499, \ k = 105.5, l = 0.0096, \ i_0 = 0.042, \ i_{fix} \approx  0.04186.$ The values of $\alpha$ are put on the left vertical axis of the bottom plot, whereas the ones of $\beta$ - on the right axis. The system starts in a recession with increased value interest rate. It decreases consequently towards the stability point, smoothly entering the phase of profitable economic growth with beneficial interest rate after some time. The cycle is closed.}
  \label{fig:cycle_recovery}
\end{figure}

The first phase of a stabilization is a recessionary period with the level of interest rate elevated by the crisis (\cref{fig:cycle_recovery}). 
The interest rate consequently decreases towards the fixed point $0.0419$ and after some time enters the area of beneficial levels, which stimulates the growth of the economy. Further in time, the pace of change significantly reduces and the value of interest rate slowly and consequently settles on the fixed point. The cycle is closed.

\subsection{Further considerations regarding the credit cycle description}

It is worth noting that the cycle was explained only by the dynamics of the financial system, without referring to external shocks interpreted as changes in $\mu, \ \nu$ parameters and utilised often as a ''divine intervention'' bringing crisis to the market. It is important because in reality the crisis can also be caused by the malfunctioning of the financial system itself \cite{krishnamurthy_amplification_2010} and modeling this phenomenon can help to avoid the crisis by proper management of the system. Naturally, external shocks expressed in the dynamics of $\mu$ and $\nu$ can also be modeled by the analyzed system \cref{eq:my_model}, providing insight into potential strategies to mitigate the crisis.

\begin{figure}[htbp]
  \centering
  \label{fig:bubble_crash_alpha}
  \includegraphics[scale=0.3]{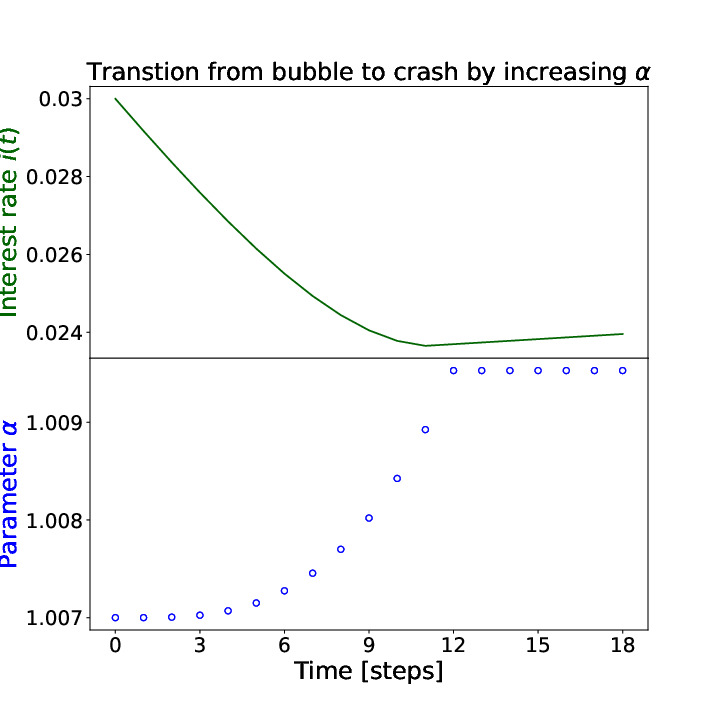}
  \caption{Transition from bubble to crash by constantly increasing $\alpha$ for $\alpha$ staring from $1.007,$  $ \beta = 1,  \ \mu = \nu = 0.499, \ k = 10^{-12}, \ l = 1.27e12, \ i_0 = 0.03.$}
  \label{fig:c_bubble_crash_alpha}
\end{figure}

The evolution of the system presented above is only one of the many possible approaches to the description of the credit cycle in the unified MWE model with dynamic parameters. It is worth to point out that it is also possible to generate the transition from a bubble to a crisis only by increasing $\alpha,$ under some parameterizations $\Lambda.$ Therefore, the analyzed model is also capable of explaining the situation where the bubble collapses because of the "excess optimism", without setting any effective constraints on the value of the interest rate (\cref{fig:c_bubble_crash_alpha}).

Parameters $\alpha$ and $\beta$ exhibit significantly different behaviors under the parameterization $\Lambda$ proposed in \cref{sec:stable_and_bubble}. They are "countermonotonic" in the sense that optimism increases in $\alpha$ and decreases in $\beta.$ On the other hand, the instability increases in both $\alpha$ and $\beta$ (as $a = \alpha \nu + \beta \nu$ describes the instability and the pace of evolution). These characteristics can lead to very 
diverse consequences near the transcritical bifurcation point $a=1,$ as various approaches to increase/decrease optimism (by modifying $\alpha$ or modifying $\beta$) can bring very different results. In the neighbourhood of the transcritical bifurcation point, more optimistic (lower) $\beta$ can help escape the unstable regime ($a<1$), although it is rather hard to expect banks to bring $\beta$ to significantly lower levels, as they should closely watch the number of defaults and express their optimism mainly in the sensitivity to the increasing amount of money in the system ($\alpha$). On the other hand, more optimism expressed in $\alpha$ drives the system to more unstable territories.
In the unstable regime, more pessimistic but careful choice of parameters (increase $\beta,$ decrease $\alpha$ such that $a<1$) can cause greater contraction in interest rate in one step (in comparison to the situation when $\alpha$ is not modified), but can also enable to avoid longer lasting crisis or recession regime.

Different possible outcomes for similar banks policies increase the uncertainty regarding the future dynamics of the system and hinder the decision-making, especially considering that there are many banks in the market and no one knows about the other's strategies. This situation is very similar to the coordination problem in bank run models \cite{diamond_bank_1983, kiss_preventing_2022}. All the described effects can amplify uncertainty and deepen panic. There is also a possibility that banks react to some minor turbulence by increasing $\alpha$ even more and hoping that more money pumped into the system will help avoid the crisis (which can be true). Nevertheless, the continuation of that strategy can lead to the situation where the growth is artificially fueled by rapidly increasing amount of money in the system, and at some point is not sustainable any more. Even if a central bank and regulations allow it, market participants will eventually switch to a stable currency to protect the value of their property \cite{bumin_predicting_2023, rochon_dollarization_2003}. 
Moving the crisis further in time can make it more severe \cite{levy_microscopic_1995, solomon_minsky_2013}.

\section{Summary}
\label{sec:summary}

In this paper, the Marshall-Walras equilibrium approach with the power law dynamics of the credit market was unified, to provide a comprehensive model of the credit cycle, describing all variables of interest and relations between them at each point in time. The model was enhanced to be Markovian, therefore eliminating the dependency on the arbitrary choice of the initial moment $t=0$. Detailed mathematical analysis of the unified model was performed to determine that it describes three very different economic regimes: stable state, bubble, and crisis, dependent on the values of model parameters. On the basis of these results, the measures of systemic risk were constructed as distances to critical values of the parameters, for which the transition between different model regimes occurs. The developed theory was applied to generate the interest rate evolution with features characteristic for a full credit cycle. For this purpose, the dynamics of the parameters was utilized in the model, with the economic interpretation of casual relationships between them. The mathematical consequences of the model and their correspondence to the real-world phenomena were analyzed in detail.

The result of this article is the unified framework for modeling credit cycles with systemic risk assessment. Our model describes various states of the market and transitions between them. The relative mathematical simplicity causes that the model can be easily operated on and incorporated into the analyses and decision-making processes performed with the use of the classical financial models. It also ensures that the model is clear and tracable, and therefore has natural economic interpretation of parameters and provides explanatory power regarding causes and effects of key market events. The standalone risk measures constructed in the article can provide information on the current market regime and possible dangers related to transitions to bubbles or crises. For example, our model can provide indicators of the presence of stable market situation with small volatility and help predict potential regime switches to unstable periods with greater variance. On the basis of this information, a financial instrument valuation or an investment strategy can be modified to prepare for a potential crisis. Other potential topics for future research include improving the model with the addition of inflation or several banking entities with interactions between them derived from empirical data. 

Research on the topic of systemic risk and credit cycles should be continued despite all the challenges connected with modeling very rare events. Complex systems, positive feedback loops, and critical points are promising tools for modeling the area of interest. Therefore, it is critical to enhance the knowledge about these phenomena, since financial systems constitute the basis of modern economies, governments, and societies, which can struggle to function properly without the foundation of a stable and secure financial system \cite{cibils_argentina_nodate}.


\bibliographystyle{siamplain}
\bibliography{Article}
\end{document}

%% file: ex_shared.tex
\usepackage{lipsum}
\usepackage{amsfonts}
\usepackage{graphicx}
\usepackage{epstopdf}
\usepackage{algorithmic}
\ifpdf
  \DeclareGraphicsExtensions{.eps,.pdf,.png,.jpg}
\else
  \DeclareGraphicsExtensions{.eps}
\fi

\newsiamremark{remark}{Remark}
\newsiamremark{hypothesis}{Hypothesis}
\crefname{hypothesis}{Hypothesis}{Hypotheses}
\newsiamthm{claim}{Claim}

\headers{The unified framework for systemic risk assessment}{K. Fortuna and J. Szwabi\'{n}ski}

\title{The unified framework for modelling credit cycles with Marshall-Walras price formation process and systemic risk assessment\thanks{Submitted 10.05.2023.
\funding{This work was supported by the Polish Ministry  of Education and Science (MEiN) core funding for statutory R\&D activities.}}}

\author{Kamil Fortuna\thanks{Faculty of Pure and Applied Mathematics, Hugo Steinhaus Center, Wrocław University of Science and Technology, Wrocław, Poland 
  (\email{kamil.fortuna@pwr.edu.pl}, \email{janusz.szwabinski@pwr.edu.pl}).}
\and Janusz Szwabi\'{n}ski\footnotemark[2]}

\usepackage{amsopn}

\makeatletter
\newcommand*{\addFileDependency}[1]{
  \typeout{(#1)}
  \@addtofilelist{#1}
  \IfFileExists{#1}{}{\typeout{No file #1.}}
}
\makeatother

\newcommand*{\myexternaldocument}[1]{%
    \externaldocument{#1}%
    \addFileDependency{#1.tex}%
    \addFileDependency{#1.aux}%
}